\documentclass[12pt]{article}

\begin{document}

\begin{center}
{\Large\bf{}The Bel-Robinson tensor and the classical pseudotensors}\\
\end{center}

\begin{center}
Lau Loi So  
\end{center}

\begin{abstract}
Describing the gravitational energy and momentum, the Bel-Robinson
tensor is the best tensor. However, the classical pseudotensors
can also manage the the same job. As Deser mentioned in 1999, a
certain linear combination between Einstein pseudotensor and
Landau-Lifshitz pseudoetnsor give a pure Bel-Robinson tensor. Here
we used the same idea but adapted the harmonic gauge, we found
that all the classical pseudotensor alone cannot give a multiple
of the Bel-Robinson tensor. But under a modification, all of them
contribute the same energy-momentum and stress as the Bel-Robinson
tensor does.
\end{abstract}

\section{Introduction}
The Bel-Robinson tensor is the best tensor to describe the
energy-momentum and stress in general relativity. Senovilla had
written a very nice article about this tensor~\cite{Senovilla}.
This tensor has many very nice mathematical properties: completely
symmetric, trace free and divergence free. In 1999,
Deser~\cite{Deser} explained how to make a certain linear
combination between Einstein~\cite{Einstein} pseudotensor and
Landau-Lifshitz~\cite{LL} pseudotensor to produce the pure
Bel-Robinson tensor, under the Riemann normal coordinates systems.
In fact, we had examined that all the classical pseudotensor alone
cannot generate a multiple of the Bel-Robinson
tensor~\cite{CQGSoNesterChen2009}. Our present paper used the same
idea, but using another angle, apply the harmonic gauge and using
the physical detail of the metric. As expected, we found none of
the classical pseudotensor can have a multiple of the pure
Bel-Robinson tensor. But after adding some modification, the
results indicate that each classical pseudotensor does have the
pure Bel-Robinson tensor, restricted in the energy-momentum and
stress. When we say the classical pseudotensors which means
Einstein, Landau-Lifshitz, Bergmann-Thomson~\cite{BT},
Papapetrou~\cite{Papapetrou} and Weinberg~\cite{Weinberg}.

\section{Technical background}

Throughout this work, we use the same spacetime signature and
notation in MTW~\cite{MTW}, including the geometrical units
$G=c=1$, where $G$ and $c$ are the Newtonian constant and the
speed of light. The Greek letters denote the spacetime and Latin
letters refer to spatial. In vacuum, the Bel-Robinson tensor can
be defined as follows
\begin{eqnarray}
B_{\alpha\beta\mu\nu}:=R_{\alpha\lambda\mu\sigma}R_{\beta}{}^{\lambda}{}_{\nu}{}^{\sigma}
+R_{\alpha\lambda\nu\sigma}R_{\beta}{}^{\lambda}{}_{\mu}{}^{\sigma}
-\frac{1}{8}g_{\alpha\beta}g_{\mu\nu}R^{2}_{\lambda\sigma\rho\tau},
\end{eqnarray}
where
$R^{2}_{\lambda\sigma\rho\tau}=R_{\lambda\sigma\rho\tau}R^{\lambda\sigma\rho\tau}$.
In order to extract the energy-momentum and the stress, one can
use the analog of the ``electric" $E_{ab}$ and ``magnetic"
$B_{ab}$ parts of the Weyl tensor \cite{Carmeli},
\begin{eqnarray}
E_{ab}:=C_{0a0b}, \quad B_{ab}:=\ast{C_{0a0b}},
\end{eqnarray}
where $C_{\alpha\beta\mu\nu}$ is the Weyl conformal tensor and
$\ast{C_{\alpha\beta\mu\nu}}$ is its dual,
\begin{equation}
\ast{C_{\alpha\beta\mu\nu}}=\frac{1}{2}
\epsilon_{\alpha\beta\lambda\sigma} C^{\lambda\sigma}{}_{\mu\nu}.
\end{equation}
(Here
$\epsilon_{\alpha\beta\mu\nu}=\epsilon_{[\alpha\beta\mu\nu]}$ with
$\epsilon_{0123}=\sqrt{-g}$ is the totally anti-symmetric
Levi-Civita tensor, see \cite{MTW}, in particular Eq. 8.10 and Ex.
8.3.) In vacuum using the Riemann tensor
\begin{equation}
E_{ab}=R_{0a0b},\quad B_{ab}=*R_{0a0b}.
\end{equation}
Certain commonly occurring quadratic combinations of the Riemann
tensor components in terms of the electric $E_{ab}$ and magnetic
$B_{ab}$ parts in vacuum are
\begin{equation}
R_{0a0b}R_{0}{}^{a}{}_{0}{}^{b}=E_{ab}E^{ab},\quad
R_{0abc}R_{0}{}^{abc}=2B_{ab}B^{ab},\quad
R_{abcd}R^{abcd}=4E_{ab}E^{ab}.
\end{equation}
In particular, the Riemann squared tensor can then be written
\begin{equation}
R^{2}_{\lambda\sigma\rho\tau} =8(E_{ab}E^{ab}-B_{ab}B^{ab}).
\end{equation}
The energy-momentum and stress for the Bel-Robinson tensor are
\begin{eqnarray}
B_{0000}&=&E^{2}_{ab}+B^{2}_{ab},\\
B_{000j}&=&2\epsilon_{jab}E^{ac}B^{b}{}_{c},\\
B_{00ij}&=&\eta_{ij}(E^{2}_{ab}+B^{2}_{ab})-2(E^{c}{}_{i}E_{cj}+B^{c}{}_{i}B_{cj}),
\end{eqnarray}
where $E^{2}_{ab}:=E_{ab}E^{ab}$ and similarly for $B^{2}_{ab}$.

Within a weak field the metric tensor can be decomposed as
$g_{\alpha\beta}=\eta_{\alpha\beta}+h_{\alpha\beta}$, and its
inverse is $g^{\alpha\beta}=\eta^{\alpha\beta}-h^{\alpha\beta}$.
Here we only consider the lowest order (see (8) at~\cite{Zhang}),
the metric components can be written as
\begin{eqnarray}
h^{00}=-E_{ab}x^{a}x^{b},\quad{}
h^{0j}=\frac{2}{3}\epsilon^{j}{}_{pq}B^{p}{}_{l}x^{q}x^{l},\quad
h^{ij}=\eta^{ij}h^{00}.
\end{eqnarray}
The lowest order for the harmonic gauge is
\begin{eqnarray}
\Gamma^{\alpha\beta}{}_{\beta}=\partial_{\beta}\bar{h}^{\alpha\beta}=0,
\end{eqnarray}
where
$\bar{h}^{\alpha\beta}=h^{\alpha\beta}-\frac{1}{2}\eta^{\alpha\beta}h$.
Thus, the energy-momentum and the stress for the Bel-Robinson
tensor can be written as the following form
\begin{eqnarray}
B^{0000}&=&\frac{1}{4}\vec{\nabla}^{2}\left[\frac{\alpha}{2}-\frac{A}{2}+\frac{B}{2}\right],\label{27aDec2022}\\
B^{000j}&=&\frac{1}{4}\vec{\nabla}^{2}(C-D),\label{27bDec2022}\\
B^{00ij}&=&\frac{1}{4}\vec{\nabla}^{2}\left[-X-Y+Z-U-V
+\eta^{ij}\left(\frac{\alpha}{2}+\frac{A}{2}-\frac{B}{2}\right)\right],\label{27cDec2022}
\end{eqnarray}
where
\begin{eqnarray}
&&\alpha=(\partial_{c}h_{00})(\partial^{c}h_{00}),
\quad{}A=(\partial_{c}h_{0d})(\partial^{c}h^{0d}),
\quad{}B=(\partial_{c}h_{0d})(\partial^{d}h^{0c}),\\
&&C=(\partial_{c}h_{00})(\partial^{c}h^{0j}),
\quad{}D=(\partial_{c}h_{00})(\partial^{j}h^{0c}),\\
&&X=(\partial^{k}h_{00})(\partial^{f}h_{00}),\quad{}
Y=(\partial^{k}h^{0c})(\partial^{f}h_{0c}),\quad{}Z=(\partial_{c}h^{0k})(\partial^{c}h^{0f}),\\
&&U=(\partial^{k}h^{0c})(\partial_{c}h^{0f}),\quad{}V=(\partial_{c}h^{0k})(\partial^{f}h^{0c}).
\end{eqnarray}

On the other hand, generally speaking, regarding the classical
pseudotensor~\cite{CQGSoNesterChen2009} which can be obtained from
a rearrangement of the Einstein field equation:
$G^{\mu\nu}=\kappa{}T^{\mu\nu}$, where the constant
$\kappa=8\pi{}G/c^{4}$, $G^{\mu\nu}$ and $T^{\mu\nu}$ are the
Einstein and stress tensors. We define the gravitational
energy-momentum density pseudotensor in terms of a suitable
superpotential $U_{\alpha}{}^{[\mu\nu]}$:
\begin{eqnarray}
2\kappa\sqrt{-g}\,t_{\alpha}{}^{\mu}:=\partial_{\nu}U_{\alpha}{}^{[\mu\nu]}-2\sqrt{-g}\,G_{\alpha}{}^{\mu}.
\label{6aMar2015}
\end{eqnarray}
Alternatively, one can rewrite (\ref{6aMar2015}) as
$\partial_{\nu}U_{\alpha}{}^{[\mu\nu]}
=\sqrt{-g}(2G_{\alpha}{}^{\mu}+2\kappa{}t_{\alpha}{}^{\mu})$.
Using the Einstein equation, the total energy-momentum density
complex can be defined as
\begin{eqnarray}
{\cal{T}}_{\alpha}{}^{\mu}:=\sqrt{-g}(T_{\alpha}{}^{\mu}+t_{\alpha}{}^{\mu})
=(2\kappa)^{-1}\partial_{\nu}U_{\alpha}{}^{[\mu\nu]}.
\end{eqnarray}
This total energy-momentum density is automatically conserved as
$\partial_{\mu}{\cal{T}}_{\alpha}{}^{\mu}\equiv{}0$, which can be
split into two parts:
\begin{eqnarray}
(\partial_{\nu}U_{\alpha}{}^{[\mu\nu]})_{1}=2\sqrt{-g}\,G_{\alpha}{}^{\mu},\quad{}
(\partial_{\nu}U_{\alpha}{}^{[\mu\nu]})_{2}=2\sqrt{-g}\kappa\,t_{\alpha}{}^{\mu}.\label{9bMar2015}
\end{eqnarray}
The first part indicates the contribution of matter and the second
piece refers to vacuum gravity~\cite{MTW}. For the equivalence
principle small limit at a point within matter, it is known that
the Einstin, Landau-Lifshitz, Bergmann-Thomson, Papapetrou and
Weinberg pseudotensors all satisfy the standard result
$2G_{\alpha}{}^{\mu}$ at the origin in Riemann normal coordinates,
while $t_{\alpha}{}^{\mu}$ vanishes in this
limit~\cite{CQGSoNesterChen2009}.

Actually there are three kinds of superpotentials:
$U_{\alpha}{}^{[\mu\nu]}$, $U^{\alpha[\mu\nu]}$ and
$H^{[\alpha\beta][\mu\nu]}$. The latter also must satisfy some
further symmetries, under the interchange of the pairs, and the
total antisymmetry of 3 indices (i.e., the same as the symmetries
of the Riemann tensor). From them the total energy-momentum
complex can be obtained as follows:
\begin{eqnarray}
2\kappa{\cal{T}}_{\alpha}{}^{\mu}=\partial_{\nu}U_{\alpha}{}^{[\mu\nu]},\quad{}
2\kappa{\cal{T}}^{\alpha\mu}=\partial_{\nu}U^{\alpha[\mu\nu]},\quad{}
2\kappa{\cal{T}}^{\alpha\mu}=\partial^{2}_{\beta\nu}H^{[\alpha\beta][\mu\nu]}.
\end{eqnarray}
In vacuum, $t^{00}$ is the gravitational energy density, $t^{0j}$
is the momentum and $t^{ij}$ is the stress. In short: we use the
following for the representation
\begin{eqnarray}
t^{\alpha\mu}=(t^{00},~t^{0j},~t^{ij})
\end{eqnarray}

\section{The classical pseudotensors}

\subsection{Einstein pseudotensor}
The following is the Freud (F) superpotential~\cite{Einstein}:
\begin{eqnarray}
_{F}U_{\alpha}{}^{[\mu\nu]}:=-\sqrt{-g}g^{\beta\sigma}\Gamma^{\tau}{}_{\lambda\beta}
\delta^{\lambda\mu\nu}_{\tau\sigma\alpha}.
\end{eqnarray}
The corresponding Einstein pseudotensor is
\begin{eqnarray}
t^{\alpha\mu}_{E}
&=&-2\Gamma^{\beta\lambda\alpha}\Gamma^{\mu}{}_{\beta\lambda}
+\Gamma^{\mu\alpha\lambda}\Gamma^{\beta}{}_{\beta\lambda}
+\Gamma^{\beta\alpha}{}_{\beta}\Gamma^{\mu\lambda}{}_{\lambda}\nonumber\\
&&+\eta^{\alpha\mu}(\Gamma^{\beta\lambda}{}_{\nu}\Gamma^{\nu}{}_{\beta\lambda}
-\Gamma^{\beta}{}_{\beta\nu}\Gamma^{\nu\lambda}{}_{\lambda})
+\Gamma^{\nu\alpha\mu}\Gamma^{\beta}{}_{\beta\nu}
-\Gamma^{\beta\alpha}{}_{\beta}\Gamma^{\lambda\mu}{}_{\lambda}.\label{15bJune2022}
\end{eqnarray}
Apply the harmonic gauge, the energy-momentum and the stress are
\begin{eqnarray}
t^{\alpha\mu}_{E}=-\left[\left(-\frac{\alpha}{2}-\frac{A}{2}+\frac{B}{2}\right),~0,~
\eta^{kf}\left(\frac{\alpha}{2}+\frac{A}{2}-\frac{B}{2}\right)-X-Y-U\right].
\end{eqnarray}
According to Table 1, we add the extra piece
\begin{eqnarray}
t^{\alpha\mu}_{\rm{New}} &=&t^{\alpha\mu}_{E}
+2t_{1}+2t_{2}-\frac{t_{3}}{2}+\frac{t_{5}}{2}-t_{9}-\frac{t_{11}}{2}\nonumber\\
&=&-\left[\frac{\alpha}{2}-\frac{A}{2}+\frac{B}{2},~C-D,~
\eta^{kf}\left(\frac{\alpha}{2}+\frac{A}{2}-\frac{B}{2}\right)-X
-Y+Z-U-V \right].
\end{eqnarray}
Then
\begin{eqnarray}
-\frac{1}{4}\vec{\nabla}^{2}t^{\alpha\mu}_{\rm{New}}=
(B^{0000},B^{000j},B^{00ij}).
\end{eqnarray}

\subsection{Landau-Lifshitz pseudotensor}
The Landau-Lifshitz (LL) superpotential~\cite{LL}
\begin{eqnarray}
U^{\alpha[\mu\nu]}_{LL}:=-(-g)g^{\alpha\beta}g^{\pi\sigma}
\Gamma^{\tau}{}_{\lambda\pi}\delta^{\lambda\mu\nu}_{\tau\sigma\beta},
\end{eqnarray}
and the pseudotensor becomes
\begin{eqnarray}
t^{\alpha\mu}_{LL}
&=&g^{\alpha\mu}(\Gamma^{\gamma}{}_{\gamma\nu}\Gamma^{\beta\nu}{}_{\beta}
+\Gamma^{\nu\beta}{}_{\rho}\Gamma^{\rho}{}_{\nu\beta}
-2\Gamma^{\gamma}{}_{\gamma\nu}\Gamma^{\nu\beta}{}_{\beta})
+2\Gamma^{\nu\alpha\mu}\Gamma^{\beta}{}_{\beta\nu}
-\Gamma^{\rho\alpha}{}_{\rho}\Gamma^{\beta\mu}{}_{\beta}\nonumber\\
&&+\Gamma^{\alpha\rho}{}_{\beta}\Gamma^{\mu\beta}{}_{\rho}
-\Gamma^{\rho\alpha}{}_{\beta}\Gamma^{\beta\mu}{}_{\rho}
-\Gamma^{\alpha\rho}{}_{\beta}\Gamma^{\beta\mu}{}_{\rho}
-\Gamma^{\rho\alpha}{}_{\beta}\Gamma^{\mu\beta}{}_{\rho}
-(\Gamma^{\alpha\mu}{}_{\nu}+\Gamma^{\mu\alpha}{}_{\nu})\,\Gamma^{\beta\nu}{}_{\beta}
\nonumber\\
&&+\Gamma^{\alpha\rho}{}_{\rho}\Gamma^{\beta\mu}{}_{\beta}
+\Gamma^{\rho\alpha}{}_{\rho}\Gamma^{\mu\beta}{}_{\beta}
-\Gamma^{\alpha\rho}{}_{\rho}\Gamma^{\mu\beta}{}_{\beta}
+(\Gamma^{\alpha\mu}{}_{\rho}+\Gamma^{\mu\alpha}{}_{\rho})\,\Gamma^{\rho\beta}{}_{\beta},\label{8aAugust2022}
\end{eqnarray}
Apply the harmonic gauge, we have the following results
\begin{eqnarray}
t^{\alpha\mu}_{LL}=-\left[\frac{7\alpha}{2}+\frac{A}{2}+\frac{B}{2},~
2C-2D,~\eta^{kf}\left(\frac{\alpha}{2}+\frac{A}{2}-\frac{B}{2}\right)-X-Y+Z-U-V\right].
\end{eqnarray}
This result is not good, then consider the modification
\begin{eqnarray}
t^{\alpha\mu}_{\rm{New}}&=&t^{\alpha\beta}_{LL} +2t_{1}+2t_{2}
-t_{3}+\frac{3}{2}t_{5} -t_{11}
\nonumber\\
&=&-\left[\frac{\alpha}{2}-\frac{A}{2}+\frac{B}{2},~
C-D,~\eta^{kf}\left(\frac{\alpha}{2}+\frac{A}{2}-\frac{B}{2}\right)-X-Y+Z-U-V\right].
\end{eqnarray}
Then we have the expected result
\begin{eqnarray}
-\frac{1}{4}\vec{\nabla}^{2}t^{\alpha\mu}_{\rm{New}}=
(B^{0000},B^{000j},B^{00ij}).
\end{eqnarray}

\subsection{Bergmann-Thomson pseudotensor}
The Bergmann-Thomson (BT) superpotential~\cite{BT} is defined as
\begin{eqnarray}
U^{\alpha[\mu\nu]}_{BT}:=-\sqrt{-g}g^{\alpha\beta}g^{\pi\sigma}
\Gamma^{\tau}{}_{\lambda\pi}\delta^{\lambda\mu\nu}_{\tau\sigma\beta},
\end{eqnarray}
and the associated pseudotensor is
\begin{eqnarray}
t^{\alpha\mu}_{BT}&=&g^{\alpha\mu}\Gamma^{\beta\lambda}{}_{\rho}\Gamma^{\rho}{}_{\beta\lambda}
-(g^{\alpha\mu}\Gamma^{\beta}{}_{\beta\rho}
-\Gamma^{\alpha\mu}{}_{\rho}-\Gamma^{\mu\alpha}{}_{\rho})\Gamma^{\rho\nu}{}_{\nu}
+\Gamma^{\alpha\nu}{}_{\nu}(\Gamma^{\beta\mu}{}_{\beta}-\Gamma^{\mu\beta}{}_{\beta})\nonumber\\
&&+(\Gamma^{\rho\alpha\mu}-\Gamma^{\alpha\mu\rho})\Gamma^{\beta}{}_{\beta\rho}
+\Gamma^{\alpha\rho}{}_{\beta}(\Gamma^{\mu\beta}{}_{\rho}-\Gamma^{\beta\mu}{}_{\rho})
-\Gamma^{\rho\alpha}{}_{\beta}(\Gamma^{\mu\beta}{}_{\rho}+\Gamma^{\beta\mu}{}_{\rho}).\label{31bDec2019}
\end{eqnarray}
Apply the harmonic gauge, the results are
\begin{eqnarray}
t^{\alpha\mu}_{\rm{New}}
=-\left[\frac{3\alpha}{2}+\frac{A}{2}+\frac{B}{2},C-D,
\eta^{kf}\left(\frac{\alpha}{2}+\frac{A}{2}-\frac{B}{2}\right)
-X-Y+Z-U-V \right].
\end{eqnarray}
This result requires some modification
\begin{eqnarray}
t^{\alpha\mu}_{\rm{New}}&=&t^{\alpha\mu}_{BT} +2t_{1}+2t_{2}
-\frac{t_{3}}{2}+\frac{3}{2}t_{5} -\frac{t_{11}}{2}
\nonumber\\
&=&-\left[\frac{\alpha}{2}-\frac{A}{2}+\frac{B}{2},C-D,
\eta^{kf}\left(\frac{\alpha}{2}+\frac{A}{2}-\frac{B}{2}\right)
-X-Y+Z-U-V \right].
\end{eqnarray}
Then we have the expected result
\begin{eqnarray}
-\frac{1}{4}\vec{\nabla}^{2}t^{\alpha\mu}_{\rm{New}}=
(B^{0000},B^{000j},B^{00ij}).
\end{eqnarray}

\subsection{Papapetrou pseudotensor}
The Papapertrou superpotential~\cite{Papapetrou} is defined as
\begin{eqnarray}
H^{[\mu\nu][\alpha\beta]}_{P}
:=-\sqrt{-g}g^{\rho\pi}\eta^{\tau\gamma}\delta_{\pi\gamma}^{\nu\mu}\delta_{\rho\tau}^{\alpha\beta},
\end{eqnarray}
equivalently one can use
$U^{\alpha[\mu\nu]}_{P}\equiv\partial_{\beta}H^{[\mu\nu][\alpha\beta]}$.
The corresponding pseudotensor is
\begin{eqnarray}
t^{\alpha\mu}_{P}&=&\eta^{\alpha\mu}(\Gamma^{\beta\lambda}{}_{\rho}\Gamma^{\rho}{}_{\beta\lambda}
+\Gamma^{\nu\beta}{}_{\nu}\Gamma^{\lambda}{}_{\lambda\beta}
-\Gamma^{\beta}{}_{\beta\rho}\Gamma^{\rho\lambda}{}_{\lambda}
+h^{\beta\nu}\Gamma^{\lambda}{}_{\lambda\beta,\nu})
\nonumber\\
&&+2\Gamma^{\alpha}{}_{\beta\lambda}\Gamma^{\mu\beta\lambda}
+\Gamma^{\alpha\beta}{}_{\beta}\Gamma^{\lambda\mu}{}_{\lambda}
+\Gamma^{\beta\alpha}{}_{\beta}\Gamma^{\mu\lambda}{}_{\lambda}
-(\Gamma^{\alpha\mu}{}_{\beta}+\Gamma^{\mu\alpha}{}_{\beta})
\,(2\Gamma^{\lambda\beta}{}_{\lambda}+\Gamma^{\beta\lambda}{}_{\lambda})
\nonumber\\
&&+h^{\alpha\beta}\partial_{\beta}\Gamma^{\mu\lambda}{}_{\lambda}
+h^{\beta\mu}\partial_{\beta}\Gamma^{\alpha\lambda}{}_{\lambda}
-h^{\alpha\mu}\partial_{\beta}\Gamma^{\beta\lambda}{}_{\lambda}
-h^{\beta\lambda}(\Gamma^{\alpha\mu}{}_{\beta,\lambda}
+\Gamma^{\mu\alpha}{}_{\beta,\lambda}).
\end{eqnarray}
Apply the harmonic gauge, we have the following results
\begin{eqnarray}
t^{\alpha\mu}_{P}
=-\left[\frac{7\alpha}{2}+\frac{A}{2}+\frac{3B}{2},~2C-2D,~
\eta^{kf}\left(\frac{\alpha}{2}+\frac{A}{2}-\frac{B}{2}\right)
-X-Y+Z-U-V \right].
\end{eqnarray}
Adding some extra pieces
\begin{eqnarray}
t^{\alpha\mu}_{\rm{New}}&=&t^{\alpha\mu}_{P}+t_{1}+t_{2}-\frac{3}{2}t_{3}+\frac{3}{2}t_{5}-\frac{t_{11}}{2}\nonumber\\
&=&-\left[\frac{\alpha}{2}-\frac{A}{2}+\frac{B}{2},C-D,
\eta^{kf}\left(\frac{\alpha}{2}+\frac{A}{2}-\frac{B}{2}\right)
-X-Y+Z-U-V \right].
\end{eqnarray}
Then we have desired result
\begin{eqnarray}
-\frac{1}{4}\vec{\nabla}^{2}t^{\alpha\mu}_{\rm{New}}=
(B^{0000},B^{000j},B^{00ij}).
\end{eqnarray}

\subsection{Weinberg pseudotensor}
The superpotential for Weinberg(W)~\cite{Weinberg} is
\begin{eqnarray}
H^{[\mu\nu][\alpha\beta]}_{W}
:=\sqrt{-\eta}\,\eta^{\alpha\pi}\eta^{\beta\xi}\eta^{\lambda\kappa}\delta^{\sigma\mu\nu}_{\pi\xi\kappa}g_{\lambda\sigma}.
\end{eqnarray}
The associated Weinberg pseudotensor becomes
\begin{eqnarray}
t^{\alpha\mu}_{W}=(2\Gamma^{\beta\alpha\mu}
-g^{\alpha\mu}\Gamma^{\beta\pi}{}_{\pi})\Gamma_{\beta}{}^{\nu}{}_{\nu}
+g^{\alpha\mu}\Gamma^{\rho\beta\nu}\Gamma_{\rho\beta\nu}
-2\Gamma^{\beta\nu\alpha}\Gamma_{\beta\nu}{}^{\mu}
-2h^{\beta\lambda}(\partial_{\lambda}\Gamma^{\alpha\mu}{}_{\beta}
-\partial^{\mu}\Gamma^{\alpha}{}_{\beta\lambda}).\label{14aOct2022}
\end{eqnarray}
Apply the harmonic gauge, we have the following results
\begin{eqnarray}
t^{\alpha\mu}_{W}
=-\left[\left(\frac{3\alpha}{2}+\frac{A}{2}+\frac{B}{2}\right),~D,~
\eta^{kf}\left(-\frac{3\alpha}{2}-\frac{3A}{2}+\frac{B}{2}\right)+X+Y-Z\right].
\end{eqnarray}
Adding some modification
\begin{eqnarray}
t^{\alpha\mu}_{\rm{New}}&=&t^{\alpha\mu}_{W}
+2t_{1}+t_{2}-\frac{t_{3}}{2}+\frac{5}{2}t_{5}-t_{6}-t_{8}-t_{9}-t_{10}+t_{11}\nonumber\\
&=&-\left[\frac{\alpha}{2}-\frac{A}{2}+\frac{B}{2},C-D,
\eta^{kf}\left(\frac{\alpha}{2}+\frac{A}{2}-\frac{B}{2}\right)-X-Y+Z-U-V\right].
\end{eqnarray}
Then we have desired result
\begin{eqnarray}
-\frac{1}{4}\vec{\nabla}^{2}t^{\alpha\mu}_{\rm{New}}=
(B^{0000},B^{000j},B^{00ij}).
\end{eqnarray}

\begin{table}[ht]
\caption{Extra superpotential for $U_{i}$.} \centering
\begin{tabular}{ccc}
\hline\hline Superpotential \\
[0.5ex] \hline
$U_{1}=h^{\mu\pi}\Gamma^{\alpha\nu}{}_{\pi}-(\mu\leftrightarrow\nu)$,&
$U_{2}=h^{\mu\pi}\Gamma^{\nu\alpha}{}_{\pi}-(\mu\leftrightarrow\nu)$,&
$U_{3}=g^{\alpha\mu}h^{\beta\nu}\Gamma^{\lambda}{}_{\lambda\beta}-(\mu\leftrightarrow\nu)
$,\\
$U_{4}=h^{\alpha\nu}\Gamma^{\mu\lambda}{}_{\lambda}-(\mu\leftrightarrow\nu)$,&
$U_{5}=h^{\alpha\mu}\Gamma^{\lambda\nu}{}_{\lambda}-(\mu\leftrightarrow\nu)$,&
$U_{6}=g^{\alpha\mu}h^{\beta\lambda}\Gamma_{\beta\lambda}{}^{\nu}-(\mu\leftrightarrow\nu)$,\\
$U_{7}=g^{\alpha\nu}h^{\beta\mu}\Gamma_{\beta}{}^{\lambda}{}_{\lambda}-(\mu\leftrightarrow\nu)
$,&
$U_{8}=g^{\alpha\nu}h^{\beta\lambda}\Gamma^{\mu}{}_{\beta\lambda}-(\mu\leftrightarrow\nu)$,&
$U_{9}=h^{\alpha\lambda}\Gamma^{\nu\mu}{}_{\lambda}-(\mu\leftrightarrow\nu)$,\\
$U_{10}=h^{\nu\pi}\Gamma_{\pi}{}^{\alpha\mu}-(\mu\leftrightarrow\nu),$
& $U_{11}=h\Gamma^{\nu\mu\alpha}-(\mu\leftrightarrow\nu) $,&
$U_{12}=g^{\alpha\mu}h\Gamma^{\lambda\nu}{}_{\lambda}-(\mu\leftrightarrow\nu)$,\\
$U_{13}=g^{\alpha\mu}h\Gamma^{\nu\lambda}{}_{\lambda}-(\mu\leftrightarrow\nu)$.\\
[1ex] \hline
\end{tabular}
\end{table}

\section{Conclusion}
In 1999, Deser proposed a certain linear combination between
Einstein and Landau Lifshitz pseudotensors to generate a multiple
of the pure Bel-Robinson tensor using the Riemann normal
coordinates. Here we use a similar idea but just try to obtain the
energy-momentum and the stress, under a harmonic gauge condition.
We found that none of the classical pseudotensor can have the
expected result by itself. But after adding some modification, all
of them have the desired results.

\end{document}